\documentclass[aps,pre,reprint,superscriptaddress]{revtex4-1}

\pdfoutput=1

\usepackage{graphicx}
\usepackage{amsmath}
\usepackage{amssymb}
\usepackage{xcolor}
\usepackage{pdfpages}
\usepackage{pgffor}
\usepackage{CJK}

\makeatletter
\AtBeginDocument{\let\LS@rot\@undefined}
\makeatother

\begin{document}

\begin{CJK*}{UTF8}{} 

\title{Polymerization of Branched Polyetherimides: Comparison between Monte Carlo Simulation and Flory-Stockmayer Theory}

\author{Chengyuan Wen ({\CJKfamily{gbsn}温程远})}
\affiliation{Department of Physics, Center for Soft Matter and Biological Physics, and Macromolecules Innovation Institute, Virginia Polytechnic Institute and State University, Blacksburg, Virginia 24061, USA}
\author{Roy Odle}
\affiliation{SABIC 1 Lexan Lane, Mt. Vernon, IN 47620, U.S.A.}
\author{Shengfeng Cheng ({\CJKfamily{gbsn}程胜峰})}
\email{chengsf@vt.edu}
\affiliation{Department of Physics, Center for Soft Matter and Biological Physics, and Macromolecules Innovation Institute, Virginia Polytechnic Institute and State University, Blacksburg, Virginia 24061, USA}

\date{\today}

\begin{abstract}
A Monte Carlo (MC) simulation method based on the Gillespie algorithm is developed for the polymerization of branched polyetherimides from two back-bone monomers [4,4'-bisphenol A dianhydride (BPADA) and m-phenylenediamine (MPD)], a chain terminator [phthalic anhydride (PA)], and a branching agent [tris[4-(4-aminophenoxy)phenyl] ethane (TAPE)]. This polymerization involves 4 reactions that all can be reduced to a condensation reaction between an amine group and a carboxylic anhydride group. By comparing the MC simulation results to the predictions of the Flory-Stockmayer theory on the molecular weight distribution, we show that the rates of the 4 reactions in the MC simulations should be set based on the concentrations of the functional groups on the monomers involved in each reaction. Using the MC simulations, we show that the Flory-Stockmayer theory predicts the molecular weight distribution well for systems below the gel point that is set by the functionality of the branching agent but starts to fail for systems around or above the gel point. However, for all the systems the MC simulations can be used to reliably predict the molecular weight distribution. Even for a macroscopic system, a converging distribution can be quickly obtained through the MC simulations on a system of only a few hundred to a few thousand monomers but with the same molar ratios of monomers as in the macroscopic system.
\end{abstract}

\maketitle

\end{CJK*}

\section{INTRODUCTION}

The molecular weight distribution and architecture are two important characteristics of a system of polymer chains \cite{RubinsteinColby2003}. They strongly affect material properties such as dynamic moduli, fracture toughness, glass transition temperature, and viscosity \cite{RubinsteinColby2003, Nunes1982, Suneel2002}. Experimental methods for an accurate determination of molecular weight distributions are thus of great interest \cite{Suneel2002, Mead1994, Williamson2016}. Theoretically, it is also highly desirable if the molecular weight distribution of a polymer can be predicted \textit{a priori} based on the knowledge of the polymerization reaction without even synthesizing the polymer. Such a theoretical method will be a valuable tool not only useful for understanding experimental measurements but also beneficial for other theories and models aiming to predict polymer properties. For example, Nichetti and Manas-Zloczower proposed a theoretical model to predict viscosity of a polymer melt based on its molecular weight distribution that was determined by fitting the gel permeation chromatography data to statistical distribution functions \cite{Nichetti1998}. A method to quickly generate molecular weight distributions before polymers are synthesized thus might be able to advance the predictive capability of such theories.

A theory on the constitution and molecular size distribution of a step-growth polymer was proposed by Flory and Stockmayer many years ago \cite{Flory1941, Flory1941a, Flory1941b, Stockmayer1943, Stockmayer1944, Stockmayer1952} and has been frequently used to determine the gel point. Flory studied the polymerization of bifunctional monomers mixed with trifunctional and tetra functional branching units and made two fundamental assumptions. First, the same functional group has the same probability to react with another group and this probability is not affected by the length of the polymer to which the functional group belongs as well as the position of the functional group on that polymer. Secondly, ring polymers are not formed. Stockmayer extended the theory to branching units with arbitrary functionalities and derived the Stockmayer formula for the number of a polymer with a given composition, though ring structures were still excluded. The predictions of the Flory-Stockmayer theory, including the gel point and the average molecular weight, have been tested experimentally \cite{Peebles1971, Matsumoto1995, Matsumoto1999, Bannister2006, GaoHaifeng2007, Schultz2009, Rosselgong2009}. However, the entire distribution is hard to probe experimentally and often only some average molecular weight is measured. Practically, it is also difficult to directly predict the molecular weight distribution using the Flory-Stockmayer theory because of mathematical complexity of computing the numbers of all possible molecules in a branched polymer system. Furthermore, the Flory-Stockmayer theory is only expected to be valid below the gel point. Beyond the gel point, the formation of cyclics and closed loops in network structures (i.e., branching) becomes significant and the theory may fail \cite{Lyu2018, Lyu2018a}.

Monte Carlo (MC) methods are a class of techniques based on random sampling to numerically solve problems that have a probabilistic interpretation \cite{Landau2005}. MC methods have broad applications in polymer science \cite{Hsu2011, Brandao2015}, especially in polymer reaction engineering \cite{Brandao2015}. Johnson and O'Driscoll used MC simulation to study sequence distributions in step-growth copolymerization \cite{Johnson1984}. Tobita applied MC simulation to a wide range of polymerization problems, including free-radical cross-linking copolymerization \cite{Tobita1993}, emulsion polymerization \cite{Tobita1995}, the modification of polymer via crosslinking and degradation \cite{Tobita1995b}, long-chain branching and random scission \cite{Tobita2001}, and living radical polymerization \cite{Tobita2006b, Tobita2007}. Hadicke and Stutz used an amine-cured epoxy as an example to compare the structure of step-growth networks obtained by MC simulation to that predicted by a branching theory \cite{Hadicke2002}. He \textit{et al.} applied a MC method to simulate self-condensing vinyl polymerization in the presence of multifunctional initiators and probed the role of reactive rate constants \cite{HeXuehao2001, HeXuehao2003}. Rouault and Milchev \cite{Rouault1997} and He \textit{et al.} \cite{HeJunpo1997} performed MC simulations to study the kinetics and chain length distributions in living polymerization. Prescott used a MC model to show that chain-length dependent termination plays a significant role in living/controlled free-radical polymerization systems containing reversible transfer agents \cite{Prescott2003}. In a series of papers, Al-Harthi \textit{et al.} used dynamic MC simulations to study atom-transfer radical polymerizations \cite{Al-Harthi2006, Al-Harthi2007, Al-Harthi2009, Al-Harthi2009a}. Polanowski \textit{et al.} \cite{Polanowski2010, Polanowski2011} and Bannister \textit{et al.} \cite{Bannister2009} used MC methods to study the branching and gelation in living copolymerization of vinyl and divinyl monomers. Recently, Lyu \textit{et al.} used a similar model to study the atom transfer radical and the conventional free radical polymerization of divinyl monomers and checked the applicability of the Flory-Stockmayer theory in such systems \cite{Lyu2018, Lyu2018a}. Gao \textit{et al.} used kinetic MC methods to simulate free radical copolymerization processes and discussed how to accelerate such simulations using scaling approaches \cite{GaoHanyu2015, GaoHanyu2017}. Meimaroglou \textit{et al.} proposed a MC algorithm to calculate the molecular weight distribution for linear polymers and the bivariate molecular weight-long chain branching distribution for highly branched polymers \cite{Meimaroglou2007}. They also used MC simulation to investigate the molecular, topological, and solution properties of highly branched low-density polyethylene \cite{Meimaroglou2011} and the ring-opening homopolymerization of L,L-Lactide \cite{Meimaroglou2017}. Iedema \textit{et al.} developed a MC simulation model including both branching and random scission to calculate the molecular weight and branching distributions and compared their calculations to experimental measurements on high-density polyethylene \cite{Iedema2013}. Yaghini and Iedema compared the results on low-density polyethylene from such MC simulations to the predictions of a multiradical model based on a Galerkin finite element approach \cite{Yaghini2015}.

One important application of MC simulations is to quickly compute molecular weight distributions \cite{Tobita1993, Tobita1995, Tobita1995b, Tobita2001, Tobita2006b, Tobita2007, Rouault1997, Milchev2000, HeJunpo1997, GaoZehui2015, Maafa2007, Soares2007, Hsu2011, Hamzehlou2013, Iedema2013, Yaghini2015, GaoHanyu2015, GaoHanyu2017, Lyu2018, Lyu2018a}. In MC simulations, all structures including rings and networks allowed by a polymerization reaction can be produced \cite{Fawcett1995}, no matter the system is below or beyond the gel point \cite{Tripathi2014}. Various schemes can also be implemented for the kinetics of the polymerization, which thus allows us to test the specific assumptions made by a theory. The Gillespie algorithm can be used to speed up the kinetics of a reaction \textit{in silico} and enable a reactive system to quickly reach a steady state \cite{Gillespie1977, Gillespie2007}. In this paper, we develop a MC simulation model based on the Gillespie algorithm to study the polymerization of polyetherimides (PEIs) in the presence of chain terminators and branching agents. The results from the MC simulations are used to test the Flory-Stockmayer theory including its assumption on the reaction rates.

This paper is organized as follows. In Sec.~\ref{sec:stockmayer} the Flory-Stockmayer theory is introduced, the technical challenge of computing molecular weight distributions with this theory is discussed, and an approximation method is proposed. In Sec.~\ref{sec:MC} we describe the MC model of the polymerization process of branched PEIs in detail. In Sec.~\ref{sec:results}, MC results are compared to the predictions of the Flory-Stockmayer theory. Practically, computations of molecular weight distributions can be only be executed for a small system either with the Flory-Stockmayer theory or the MC model. We thus include a discussion on the effect of finite system size in this section. Although the emphasis is on stoichiometric, fully reacted systems, those that are only partially reacted and/or nonstoichiometric are also discussed in Sec.~\ref{sec:results}. Finally, a brief summary is provided in Sec.~\ref{sec:conclusions}.

\section{Flory-Stockmayer Theory of Step-Growth Polymers} \label{sec:stockmayer}

Flory and Stockmayer considered a general step-growth polymer that consists of two types of monomers, $A$ and $B$. All reactions occur between $A$ and $B$. There are $i$ type-$A$ monomers denoted as $A_1$, $A_2$, ..., $A_i$. To simplify notation, we also use $A_q$ with $q\in \{1, 2, ..., i\}$ to denote the number of $A_q$ monomers. Similarly, there are $j$ type-$B$ monomers and the corresponding numbers are $B_1$, $B_2$, ..., $B_j$, respectively. The symbol $f_q$ denotes the functionality of an $A_q$ monomer, where $q\in \{1, 2, ..., i\}$, i.e., there are $f_q$ functional groups on an $A_q$ monomer that can form bonds with the corresponding functional groups on a $B_h$ monomer, where $h\in \{1, 2, ..., j\}$. The functionality of a $B_h$ monomer is denoted as $g_h$. The Flory-Stockmayer theory can be applied to a polymerized state where a fraction $p_A$ of all the functional groups on the type-$A$ monomers have reacted with a fraction $p_B$ of all the functional groups on the type-$B$ monomers. Therefore,
\begin{align}
\label{eq:constraint}
p_A\sum_{q=1}^i f_qA_q=p_B\sum_{h=1}^j g_hB_h.
\end{align}
In this paper, we call the systems with $\sum_{q=1}^i f_qA_q=\sum_{h=1}^j g_hB_h$ and thus $p_A = p_B$ as stoichiometric systems while those with $\sum_{q=1}^i f_qA_q \neq\sum_{h=1}^j g_hB_h$ and $p_A \neq p_B$ as nonstoichiometric. The systems with $p_A$ or $p_B$, or both, equal to 1 are fully reacted.

We use $N\{m,n\}$ to denote the number of molecules formed by $m_q$ monomers of sub-type $A_q$ and $n_h$ monomers of sub-type $B_h$, with $q$ running from 1 to $i$ and $h$ running from 1 to $j$. Here $\{m,n\}$ is a shorthand of $\{m_1, m_2., ..., m_i, n_1, n_2, ..., n_j\}$, which denotes the monomer composition of a given molecule. The Flory-Stockmayer theory predicts that
\begin{align}
\label{eq:stockmayer_dist}
\begin{split}
N\{m,n\}&=K\frac{\left(\sum_{q=1}^i f_q m_q-\sum_{q=1}^i m_q\right)!}{\left(\sum_{q=1}^i f_q m_q-\sum_{q=1}^i m_q-\sum_{h=1}^j n_h+1\right)!}\\
&\times \frac{\left(\sum_{h=1}^j g_h n_h-\sum_{h=1}^j n_h\right)!}{\left(\sum_{h=1}^j g_h n_h-\sum_{h=1}^j n_h-\sum_{q=1}^i m_q+1\right)!} \\
&\times \prod_{q=1}^i \frac{x_q^{m_q}}{m_q!}\prod_{h=1}^j \frac{y_h^{n_h}}{n_h!}
\end{split}
\end{align}
with
\begin{align}
x_q&=\frac{f_qA_q}{\sum_{l=1}^i f_lA_l}\frac{p_B\left(1-p_A\right)^{f_q-1}}{(1-p_B)}~,\\
y_h&=\frac{g_hB_h}{\sum_{l=1}^j g_lB_l}\frac{p_A\left(1-p_B\right)^{g_h-1}}{1-p_A}~,\\
K&=\frac{\left(1-p_A\right)\left(1-p_B\right)}{p_B }\sum_{q=1}^i f_qA_q \nonumber \\
&= \frac{\left(1-p_A\right)\left(1-p_B\right)}{p_A} \sum_{h=1}^jg_hB_h~.
\end{align}

Equation (\ref{eq:stockmayer_dist}) is called the Stockmayer formula, which gives the number of molecules of any monomer compositions. However, practically it is difficult to compute the molecular weight distribution from the Stockmayer formula, as all the possible combinations for $\{m_1, m_2., ..., m_i, n_1, n_2, ..., n_j\}$ have to be taken into account. Since $m_q$ runs from 1 to $A_q$ for $q\in \{1, 2, ..., i\}$ and $n_h$ runs from 1 to $B_h$ for $h\in \{1, 2, ..., j\}$, the total number of possible molecules is $\prod_{q=1}^i A_q!\times \prod_{h=1}^j B_h!$. This number is huge when there are many sub-types (i.e., large $i$ and $j$) and/or large numbers (i.e., large $A_q$ and $B_h$) of monomers involved in a polymerization.

For a molecule with composition $\{m,n\}$, the total number of monomers is $\sum_{q=1}^i m_q + \sum_{h=1}^j n_h$. Since the Flory-Stockmayer theory does not consider rings, the total number of bonds in this molecule must be $\sum_{q=1}^i m_q + \sum_{h=1}^j n_h-1$. When $p_A=p_B=1$, all the functional groups have reacted and in a given molecule the total number of the functional groups on all the type-$A$ monomers is equal to the total number of the functional groups on all the type-$B$ monomers. This number must also be equal to the total number of bonds in that molecule. Namely, for $p_A=p_B=1$ there are two identities,
\begin{align}
\sum_{q=1}^i f_q m_q=\sum_{q=1}^i m_q+\sum_{h=1}^j n_h-1
\label{eq:fully_reacted_constraint_1}
\end{align}
and
\begin{align}
\sum_{h=1}^j g_h n_h=\sum_{h=1}^j n_h+\sum_{q=1}^i m_q-1.
\label{eq:fully_reacted_constraint_2}
\end{align}
These two identities can help us simplify the Stockmayer formula for stoichiometric, fully reacted systems. Note that the terms in Eq.~(\ref{eq:stockmayer_dist}) involving $\left(1-p_A\right)$ and $\left(1-p_B\right)$ appear as
$$\left(1-p_A\right)^{\sum_{q=1}^i f_q m_q-\sum_{q=1}^i m_q-\sum_{h=1}^j n_h+1}$$
and
$$\left(1-p_B\right)^{\sum_{h=1}^j g_h n_h-\sum_{h=1}^j n_h-\sum_{q=1}^i m_q+1}~.$$
These terms can be dropped out because of Eqs.~(\ref{eq:fully_reacted_constraint_1}) and (\ref{eq:fully_reacted_constraint_2}). As a result, for fully reacted stoichiometric systems with $p_A=p_B=1$ the Stockmayer formula is simplified as
\begin{align}
\label{eq:stockmayer_dist_simp}
N\{m,n\}&=K\left(\sum_{q=1}^i f_q m_q-\sum_{q=1}^i m_q\right)!\nonumber \\
&\times \left(\sum_{h=1}^j g_h n_h-\sum_{h=1}^j n_h\right)!
\prod_{q=1}^i \frac{x_q^{m_q}}{m_q!}\prod_{h=1}^j \frac{y_h^{n_h}}{n_h!}
\end{align}
with
\begin{align}
x_q&=\frac{f_qA_q}{\sum_{l=1}^i f_lA_l}~,\\
y_h&=\frac{g_hB_h}{\sum_{l=1}^jg_lB_l}~,\\
K&=\sum_{q=1}^i f_qA_q =\sum_{h=1}^jg_hB_h~.
\end{align}

Computing $N\{m,n\}$ is not easy as it contains many factorials. The calculation can be expedited using the Stirling approximation,
\begin{align}
\label{eq:logstirling}
\log n!\approx \log\left(\sqrt{2\pi n}\right)+ n\log \left(\frac{n}{e}\right)+\log\left(1+\frac{1}{12n}\right)~.
\end{align}
Then for fully reacted stoichiometric systems, the Stockmayer formula can be approximated logarithmically as
\begin{align}
\label{eq:logN}
\log N\{m,n\} & \approx  \log K+ \log \left(\sum_{q=1}^i f_q m_q-\sum_{q=1}^i m_q\right)! \nonumber \\
& +
\log\left(\sum_{h=1}^j g_h n_h-\sum_{h=1}^j n_h\right)! \nonumber \\
&+\sum_{q=1}^i \left(m_q\log x_q -\log m_q ! \right) \nonumber \\
&+\sum_{h=1}^j \left(n_h\log y_h -\log n_h !\right)~.
\end{align}

The computation of the molecular weight distribution from $N\{m,n\}$ can be further accelerated by noting that not all the combinations $\{m,n\}$ will yield a molecule. For a fully reacted stoichiometric system where $p_A=p_B=1$, Eqs.~(\ref{eq:constraint}), (\ref{eq:fully_reacted_constraint_1}), and (\ref{eq:fully_reacted_constraint_2}) can be used to reduce the total number of $\{m,n\}$. For the branched PEIs considered in this paper (see Sec.~\ref{sec:MC}), $f_1=1$, $f_2=2$, $g_1=2$ and $g_2=3$. The constraints become
\begin{align}
\label{eq:constrain_PEI_1}
m_1+2m_2&=2n_1+3n_2~,
\end{align}
and
\begin{align}
\label{eq:constrain_PEI_2}
m_2 = n_1+n_2-1~.
\end{align}
Combining Eqs.~(\ref{eq:constrain_PEI_1}) and (\ref{eq:constrain_PEI_2}), we get 
\begin{align}
\label{eq:constrain_PEI_3}
	m_1= n_2+2~.
\end{align}

Equations (\ref{eq:constrain_PEI_2}) and (\ref{eq:constrain_PEI_3}) indicate that $m_1$ and $m_2$ are totally constrained by $n_1$ and $n_2$ in an allowed composition. Furthermore, since $m_2\geq 0$, $n_1$ and $n_2$ cannot be zero at the same time. The time complexity to enumerate all possible molecules is thus $\mathcal{O}(B_1B_2)$, which is about $\mathcal{O}(Z^2)$ with $Z$ being the system size (i.e., the total number of monomers prior to polymerization). This time complexity is acceptable for small systems. However, if there are more sub-types of monomers, then the time complexity will increase exponentially as $\mathcal{O}(Z^w)$, where $w$ is the number of monomer sub-types. For not fully reacted or nonstoichiometric systems where $p_A$ or $p_B$ are less than 1, we lose the constrains that help reduce the number of possible $\{m,n\}$ and then computing the molecular weight distribution from $N\{m,n\}$ has to rely on Eq.~(\ref{eq:stockmayer_dist}) and will become more challenging, even though the Stirling approximation may still be used. In these situations, the MC model described below will serve as a solution as it does not suffer from such limitations and the time complexity of computing the molecular weight distribution with MC simulations is always $\mathcal{O}(Z)\times k$, where $k$ is the number of MC runs needed to obtain desired statistics. Usually, $k$ is about $10^3$ to $10^4$.

\section{Monte Carlo Model of Polymerization of Branched Polyetherimides}
\label{sec:MC}

\begin{figure}[h]
\includegraphics[width = 0.48\textwidth]{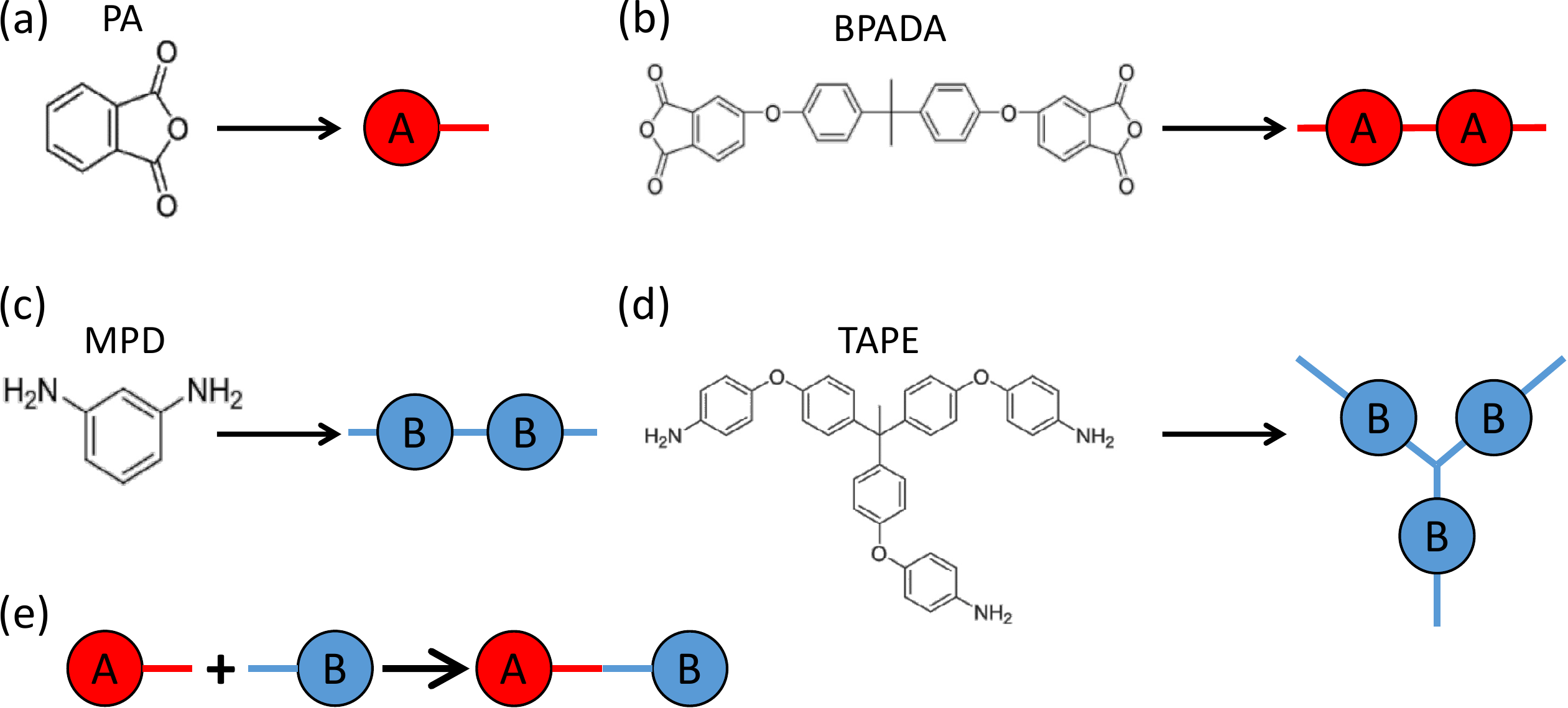}
\caption{(a)-(d): The representation of the four types of monomers of branched PEIs in the MC simulation model. Each functional group containing one amine is mapped to a $B$ bead. Each functional group containing one carboxylic anhydride is mapped to an $A$ bead. (e): Each $A$ bead can form a bond with a $B$ bead, mimicking the condensation reaction between an amine group and a carboxylic anhydride group in the polymerization of PEIs.}
\label{fig:reaction_scheme}
\end{figure}

Four types of monomers are involved in the formation of branched PEIs, including 4,4'-bisphenol A dianhydride (BPADA), m-phenylenediamine (MPD), phthalic anhydride (PA), and tris[4-(4-aminophenoxy)phenyl] ethane (TAPE).\cite{Odle2018} The chemical structures of these monomers are shown in Fig.~\ref{fig:reaction_scheme}. The involved reaction is the condensation reaction between an amine group on MPD or TAPE and a carboxylic anhydride group on BPADA or PA. In the notation of the Flory-Stockmayer theory, PA is monomer $A_1$ with $f_1 = 1$, BPADA is monomer $A_2$ with $f_2 = 2$, MPD is monomer $B_1$ with $g_1 = 2$, and TAPE is monomer $B_2$ with $g_2 = 3$. Out of these monomers, PA is an end capper to terminate a chain and TAPE is a trifunctional branching agent. Fig.~\ref{fig:reaction_scheme} shows the representation of these monomers in our MC model. Each functional group containing one carboxylic anhydride is mapped to an $A$ bead and that containing one amine is mapped to a $B$ bead. Each $A$ bead can react with a $B$ bead to form a bond (i.e., $A+B\rightarrow AB$), which describes the condensation reaction between an amine group and a carboxylic anhydride group.

\begin{figure}[h]
\includegraphics[width = 0.48\textwidth]{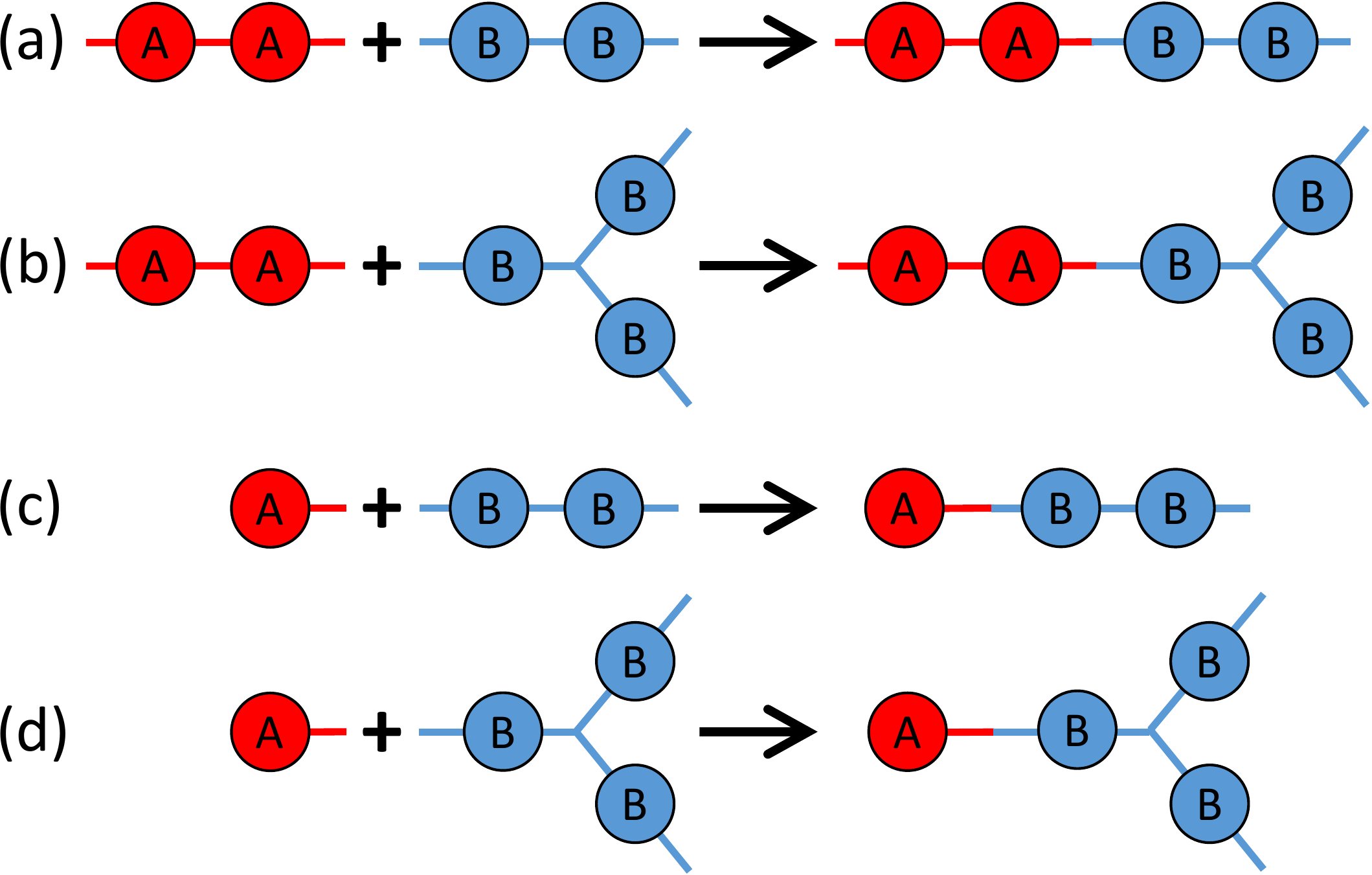}
\caption{The four reactions occurring in the polymerization of branched PEIs: (a) Reaction 1: BPADA + MPD, (b) Reaction 2: BPADA + TAPE, (c) Reaction 3: PA + MPD, and (d) Reaction 4: PA + TAPE.}
\label{fig:4_reactions}
\end{figure}

In the formation of branched PEIs consisting of the above 4 types of monomers, there are 4 possible reactions, as sketched in Fig.~\ref{fig:4_reactions}. Reaction 1 is between BPADA and MPD, which leads to the formation of a PEI backbone. Reaction 2 is between BPADA and TAPE that results in branching. Reaction 3 is between PA and MPD, which terminates a PEI chain. Reaction 4 is between PA and TAPE, which consumes one amine group on TAPE and effectively reduces its functionality by 1.

With the mapping in Fig.~\ref{fig:reaction_scheme} and the reaction scheme in Fig.~\ref{fig:4_reactions}, we perform MC simulations to study the polymerization of branched PEIs. We adopt the Gillespie algorithm to speed up MC simulations. Since only the final chain structures are concerned, we neglect the random process in the typical Gillespie algorithm that determines the time interval after which the next reaction occurs. We only keep the random process of picking a reaction at a time. At each MC step, all 4 reactions will have a probability to occur and the reaction rate of a particular reaction is determined by a rate constant and the concentration of the two types of monomers involved in that reaction. Mathematically, the probability of reaction $l$ is proportional to
\begin{align}\label{eq:reaction_rate}
P_l(L_l+R_l\rightarrow L_lR_l)=k_l n_{L_l}n_{R_l}~,
\end{align}
where $L_l$ ($R_l$) represents the reactant consisting of $A$ ($B$) beads, $k_l$ is a rate constant, $n_{L_l}$ ($n_{R_l}$) is a quantity that depends on the concentration of the reactant $L_l$ ($R_l$), and $l \in \{1, 2, 3, 4\}$ indexes the possible reactions sketched in Fig.~\ref{fig:4_reactions}. Specifically, $L_1$ and $L_2$ are BPADA, $L_3$ and $L_4$ are PA, $R_1$ and $R_3$ are MPD, and $R_2$ and $R_4$ are TAPE. Since all the 4 reactions can be reduced to the reaction between an $A$ bead and a $B$ bead (i.e., the reaction between a functional group containing one amine and another functional group containing one carboxylic anhydride) as shown in Fig.~\ref{fig:reaction_scheme}(e), $k_l$ will be set as a constant $k$ for all the 4 reactions and $n_{L_l}$ and $n_{R_l}$ can be expressed as
\begin{align}
\label{eq:concentration}
\begin{split}
	& n_{L_1}=n_{L_2}=2n_\text{BPADA}~,\\
	& n_{L_3}=n_{L_4}=n_\text{PA}~,\\
	& n_{R_1}=n_{R_3}=2n_\text{MPD}~,\\
	& n_{R_2}=n_{R_4}=3n_\text{TAPE}~,
\end{split}
\end{align}
where $n_\text{BPADA}$, $n_\text{PA}$, $n_\text{MPD}$, and $n_\text{TAPE}$ are the concentrations of monomers available for reactions (i.e., monomers with at least one unreacted functional group). In other words, $n_{L_l}$ ($n_{R_l}$) is the concentration in terms of the number of $A$ ($B$) beads on the reactant $L_l$ ($R_l$). The particular reason of writing $n_{L_l}$ and $n_{R_l}$ in this way will be discussed in Sec.~\ref{sec:rate_constant}.

\begin{figure}
\center
\includegraphics[width = 0.45\textwidth]{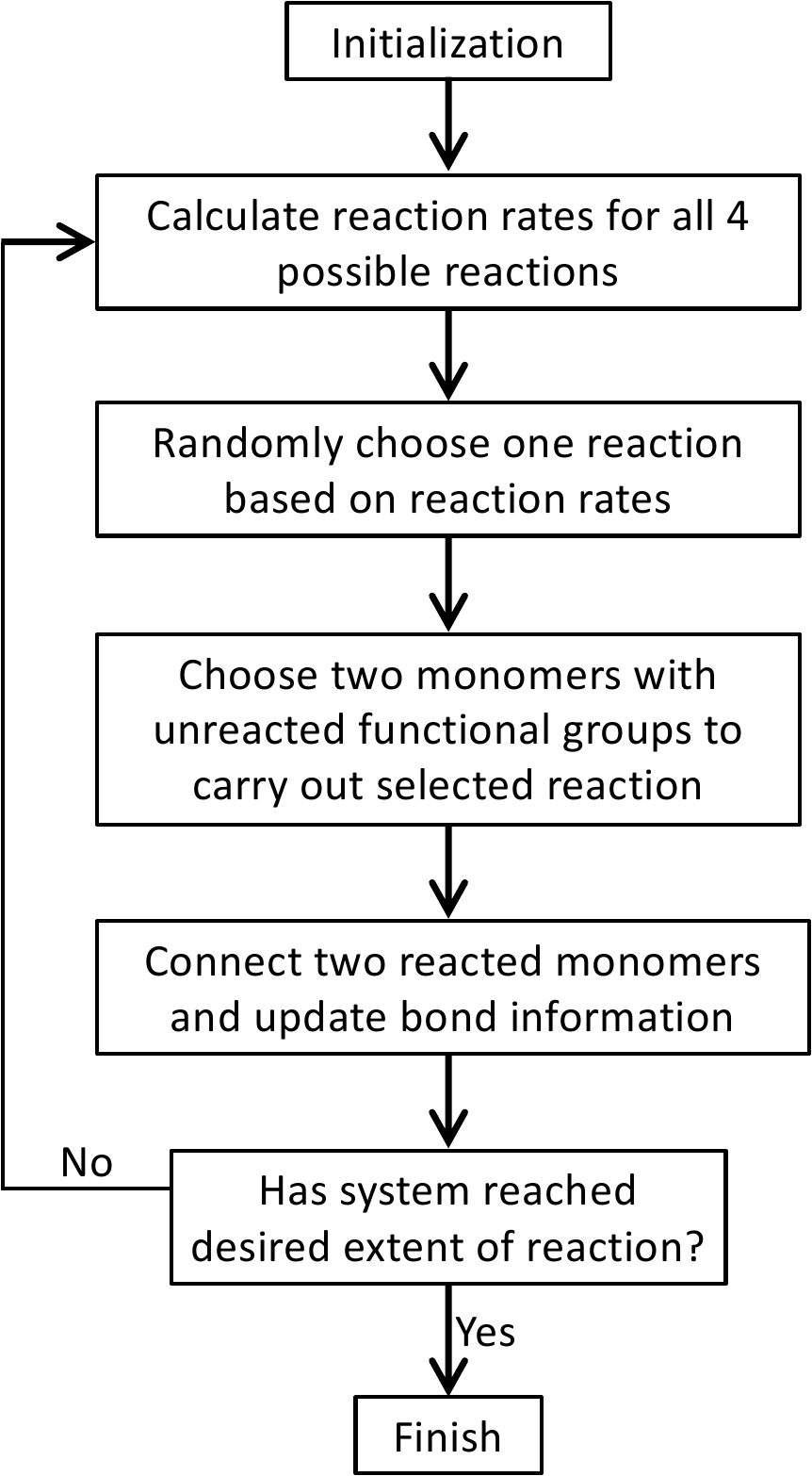}
\caption{The flow chart of the MC simulation model}
\label{fig:flow_chart}
\end{figure}

At each MC step, the probability of Reaction $l$ to be chosen is equal to $P_l/\sum_{q=1}^4 P_q$. After a reaction is selected, a pair of $L_l$ and $R_l$ that have unreacted functional groups (i.e., with unreacted $A$ and $B$ beads, respectively) is randomly chosen to react. Then the system status is updated, including the bond information between the monomers and the identity of monomers with unreacted functional groups. The MC process is repeated for the updated system until no more reactions can occur or the system has reached a desired extent of reaction. The flow chart of the MC simulation model is shown in Fig.~\ref{fig:flow_chart}. Note that in this model, we made a simplification by not allowing backward reactions, which means that once formed, the bond between an $A$ bead and a $B$ bead cannot break. However, the model permits the formation of  both rings and networks.

\section{Results and Discussion} \label{sec:results}

\subsection{Rate Constant $k$} \label{sec:rate_constant}

Equation (\ref{eq:concentration}) indicates that the reaction rate $P_l$ is based on the concentrations of the functional groups (i.e., $A$ beads or $B$ beads) on the reactants involved in that reaction. However, $P_l$ can also be computed from the concentrations of the available reactants themselves, i.e., the monomer concentrations. In this case, the reaction rate $P_l$ can be written in the same way as in Eq.~(\ref{eq:reaction_rate}) but with Eq.~(\ref{eq:concentration}) replaced by
\begin{align}
\label{eq:concentration_2}
\begin{split}
	& n_{L_1}=n_{L_2}=n_\text{BPADA}~,\\
	& n_{L_3}=n_{L_4}=n_\text{PA}~,\\
	& n_{R_1}=n_{R_3}=n_\text{MPD}~,\\
	& n_{R_2}=n_{R_4}=n_\text{TAPE}~.
\end{split}
\end{align}

To check which way of computing the reaction rates yields results that are more applicable to realistic systems, we performed a test with a simple system consisting of only PA and TAPE monomers, as shown in Table~\ref{table:test_reaction_rate}. For this system, there are only 4 possible final products, including single TAPEs and TAPEs connected with one, two, or three PAs, respectively.

\begin{table}[h]
\center
\begin{tabular}{|c|c|c|c|c|c|}
\hline
 Monomer & PA   & BPADA & MPD & TAPE \\ \hline
 Number & 2000 & 0     & 0   & 1000 \\ \hline
\end{tabular}
\caption{System used for checking the way to compute the reaction rates.}
\label{table:test_reaction_rate}
\end{table}

\begin{figure}
\center
\includegraphics[width = 0.45\textwidth]{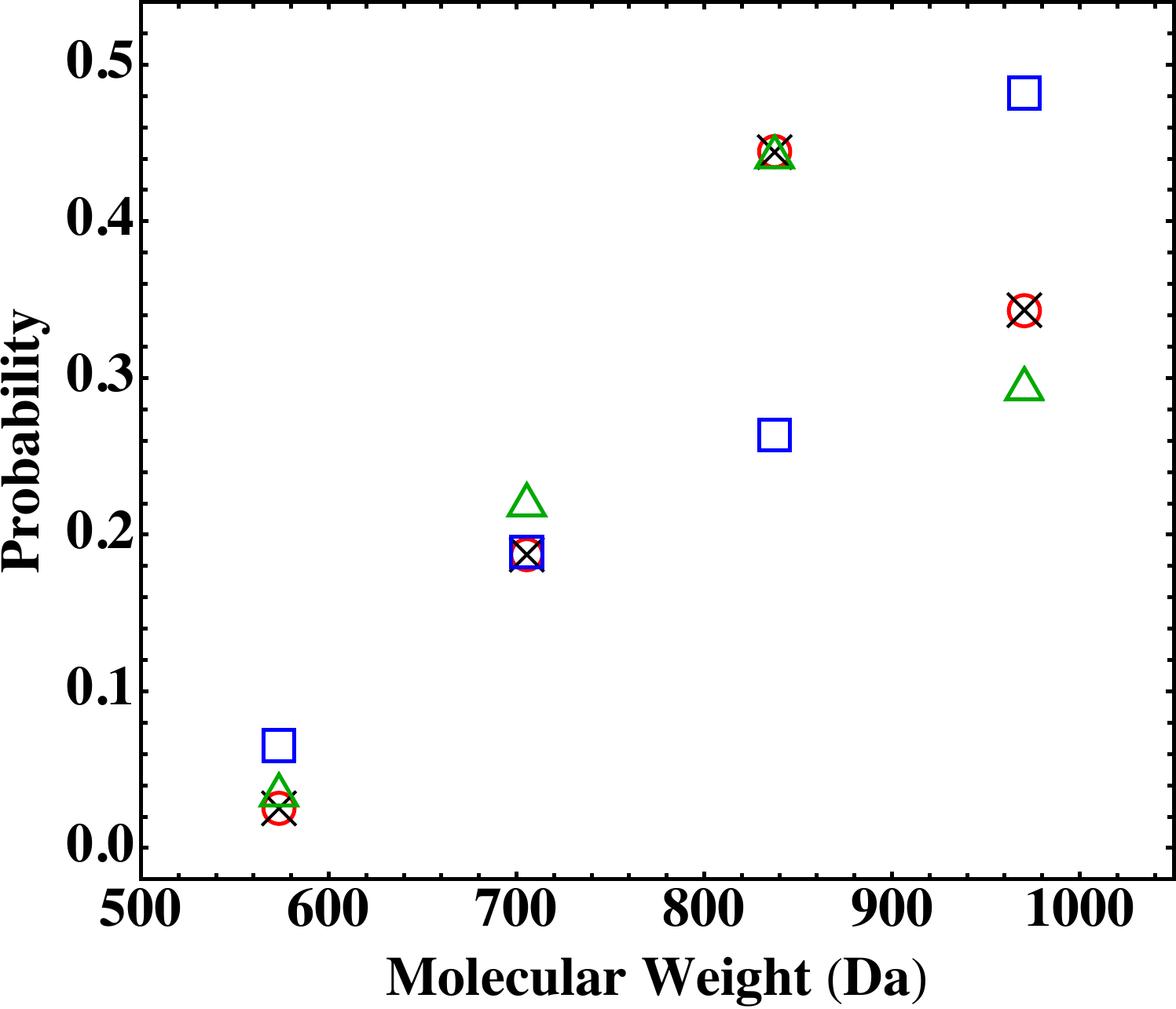}
\caption{Probabilities of the 4 possible final products for the system in Table~\ref{table:test_reaction_rate}. Results are from the Flory-Stockmayer theory (red circles), the MC simulations based on Eq.~(\ref{eq:concentration}) (black crosses), the MC simulations based on Eq.~(\ref{eq:concentration_2}) (blue squares), and a simple statistical model discussed in the main text (green triangles). The MC results are averages of 10000 runs.}
\label{fig:check_reaction_rate}
\end{figure}

Figure~\ref{fig:check_reaction_rate} shows the results on the probability distribution of the 4 final products for the system in Table~\ref{table:test_reaction_rate}, for which gelation is not a concern. The comparison shows that the results from the MC simulations based on Eq.~(\ref{eq:concentration}) agree with the Flory-Stockmayer theory while those based on Eq.~(\ref{eq:concentration_2}) do not. Furthermore, for the molar ratio in Table~\ref{table:test_reaction_rate}, all the anhydride groups on the PA monomers are reacted and each amine group on a TAPE monomer has a 2/3 chance to be reacted in a fully reacted system. A simple statistical analysis shows that the probabilities for a TAPE monomer to react with 0, 1, 2, and 3 PAs are $\left(\frac{1}{3}\right)^3$, $3\times \frac{2}{3}\times \left(\frac{1}{3}\right)^2$, $3\times \left(\frac{2}{3}\right)^2\times \frac{1}{3}$, and $\left(\frac{2}{3}\right)^3$, respectively. These results are also plotted in Fig.~\ref{fig:check_reaction_rate} and are close to those from the Flory-Stockmayer theory and the MC simulations based on Eq.~(\ref{eq:concentration}). The small differences are due to the fact that the theory and simulations consider a finite system while the statistical model assumes an infinite system. We conclude that the reactions rates based on Eq.~(\ref{eq:concentration}) should be used in the MC simulations. From now on all the data presented in this paper are computed with Eq.~(\ref{eq:concentration}) for the reaction rates. In the next two sections, Sec.~\ref{sec:full_sys} and Sec.~\ref{sec:diff_size_sys}, we focus on fully reacted stoichiometric systems where $p_A=p_B=1$. We discuss partially reacted stoichiometric systems where $p_A=p_B<1$ in Sec.~\ref{sec:nfr_sys} and nonstoichiometric systems where $p_A \neq p_B$ in Sec.~\ref{sec:neq_sys}.

\subsection{Fully Reacted Stoichiometric Systems} \label{sec:full_sys}

For the branched PEIs considered in this paper, type $A$ monomers are BPADA and PA and type $B$ monomers are MPD and TAPE, with $f_1=1$, $f_2=2$, $g_1=2$ and $g_2=3$. From the Flory theory \cite{Flory1941}, the gel point is $\alpha_c=1/(g_2-1)=1/2$. However, for the systems at hand the expression of $\alpha$, which characterizes the level of cross-linking, has to be modified because each PA monomer as a chain terminator has only one functional group. The modified expression is
\begin{align}
\alpha&=\sum_{q=0}^\infty\left[p_A(1-\rho_1)p_B(1-\rho_2)\right]^q p_A(1-\rho_1)p_B\rho_2 \nonumber \\
&=p_Ap_B\frac{(1-\rho_1)\rho_2}{1-p_Ap_B(1-\rho_1)(1-\rho_2)}~,
\label{eq:alpha}
\end{align}
where $\rho_1$ is the fraction of functional groups on the terminators (i.e., PA monomers) with respect to all the functional groups on type $A$ monomers and $\rho_2$ is the fraction of functional groups on the branching agents (i.e., TAPE monomers) with respect to all the functional groups on type $B$ monomers. For a fully reacted stoichiometric system, $p_A=p_B=1$ and this expression can be simplified as
\begin{align}
\alpha=\frac{(1-\rho_1)\rho_2}{\rho_1+\rho_2-\rho_1\rho_2}~.
\label{eq:alpha_2}
\end{align}

\begin{table*}[h]
\resizebox{\textwidth}{!}{%
\begin{tabular}{|c|c|c|c|c|c|c|c|c|c|c|c|c|}
\hline
           & PA   & BPADA & MPD & TAPE & $\rho_1$ &$\rho_2$ & $p_A$ & $p_B$ & $\alpha$ & $M_n$ (Da) & $M_w$ (Da)  & $M_z$ (Da) \\ \hline
$S_<$ & 50 & 670     & 680   & 10 &0.0360 & 0.0216 & 1 & 1 & 0.366 & $19120 \pm 33$ & $52126\pm 545$ & $78671\pm 2488$ \\ \hline
$S_\simeq$ & 50 & 670     & 671   & 16 &0.0360 & 0.0345 & 1 & 1 & 0.481 & $22000\pm 15$ & $77353\pm 307$ & $116227\pm 499$ \\ \hline
$S_>$ & 50 & 670     & 620   & 50 &0.0360 & 0.108 & 1 & 1 & 0.743 & $51055\pm 125$ & $330124\pm458$  & $369588\pm 348$ \\ \hline
\end{tabular}}
\caption{Three fully reacted, stoichiometric systems below, around, and beyond the gel point. The first column is the system label. The next 4 columns list the number of each type of monomers. The values of $\rho_1$ and $\rho_2$ are determined from the monomer numbers. The value of $\alpha$ is computed using Eq.~(\ref{eq:alpha}). The average molecular weights, $M_n$, $M_w$, and $M_z$, are from the MC simulations.}
\label{table:casestb1}
\end{table*}

We can vary the numbers of monomers to tune $\rho_1$ and $\rho_2$, thus putting the fully reacted system below, around, or beyond the gel point. Three such systems are listed in Table~\ref{table:casestb1}, where $\rho_2$ is changed by varying the numbers of MPD and TAPE monomers. In the MC simulations of these stoichiometric systems, we set $p_A = p_B = 1$, thus allowing the systems to be fully reacted.

\begin{figure}[h]
\center
\includegraphics[width = 0.45\textwidth]{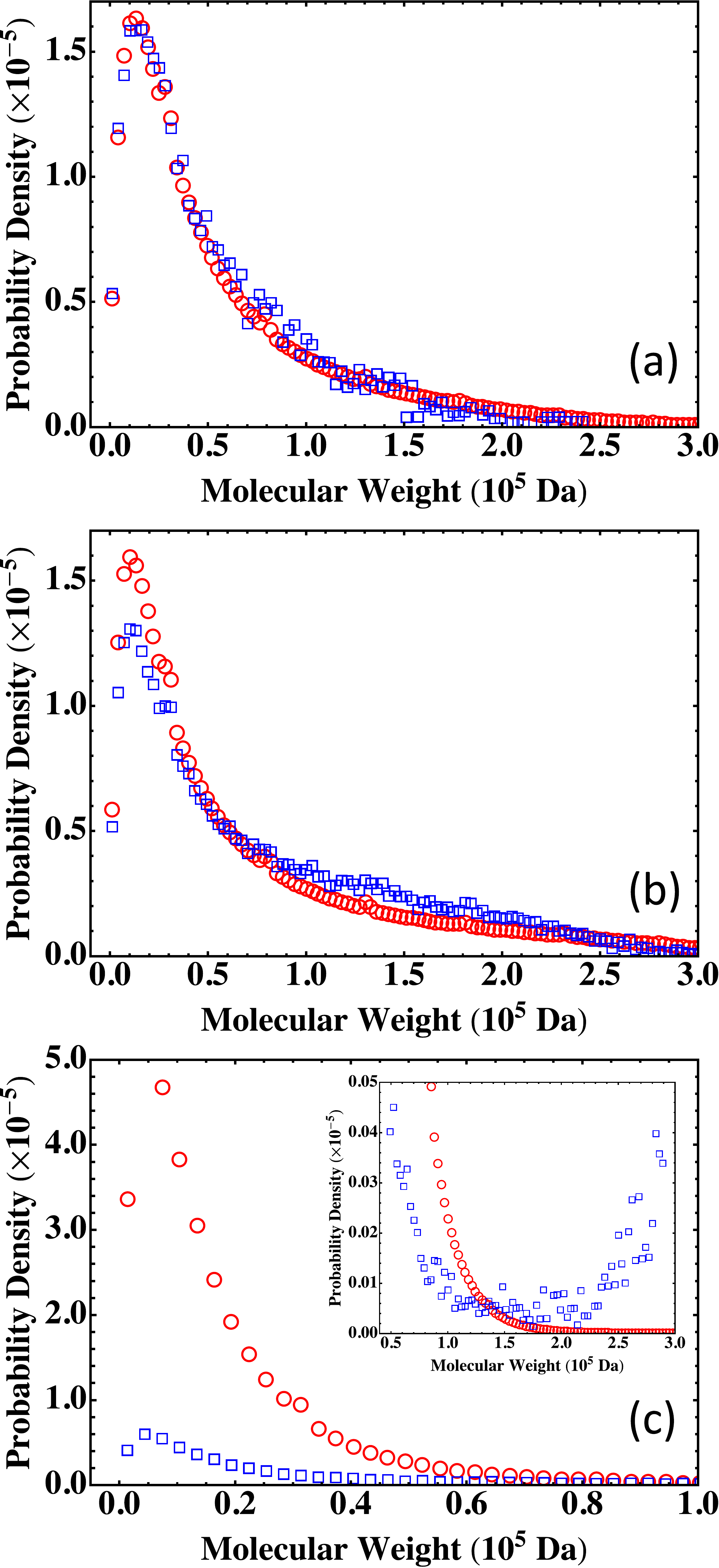}
\caption{Molecular weight distribution for the three systems in Table~\ref{table:casestb1}: (a) $S_<$, (b) $S_\simeq$, and (c) $S_>$. The results are for the Flory-Stockmayer theory (red circles) and the MC simulations (blue squares). The MC results are averages of 1000 runs for $S_<$ and 10000 runs for $S_\simeq$ and $S_>$.}
\label{fig:three_systems}
\end{figure}

The results on the molecular weight distribution from the Flory-Stockmayer theory and the MC simulations are shown in Fig.~\ref{fig:three_systems}. The comparison shows that for a system below the gel point such as $S_<$ (Fig.~\ref{fig:three_systems}(a)), the MC results agree well with the Flory-Stockmayer theory. For $S_\simeq$ which is close to the gel point, the Flory-Stockmayer theory overestimates the fraction of low molecular weight polymers and underestimates the fraction of high molecular weight species when compared to the results from the MC simulations, as shown in Fig.~\ref{fig:three_systems}(b). The discrepancy between the Flory-Stockmayer theory and the MC results becomes more dramatic for systems above the gel point. For $S_>$, $\alpha = 0.74308$, way above the critical gel point $\alpha_c = 0.5$. The Flory-Stockmayer theory predicts a probability density that is about 8 times of the MC result in the region of low molecular weight from 0 to about $0.5\times 10^5$ Da, as shown in Fig.~\ref{fig:three_systems}(c). However, the MC simulations show a significant fraction of polymers in the region of molecular weight higher than about $1.5\times 10^5$ Da and these high molecular weight polymers are completely overlooked by the Flory-Stockmayer theory, as shown in the inset of Fig.~\ref{fig:three_systems}(c). This discrepancy is not surprising as beyond the gel point, polymers with a large network structure are expected and closed loops can frequently emerge in such polymers. The Flory-Stockmayer theory does not consider the formation of rings and thus cannot accurately predict the molecular weight distribution for systems above the gel point.

\subsection{Effect of System Size}  \label{sec:diff_size_sys}

In experiments, the amount of monomers involved is at the order of moles, i.e., at the order of $10^{23}$. It is thus practically impossible to directly compute the molecular weight distribution from the Stockmayer formula (Eq.~(\ref{eq:stockmayer_dist})) for such macroscopic systems. These systems are also out of the reach of MC simulations that typically deals with systems of fewer than $10^6$ monomers. A natural question we can ask is: if we keep the molar ratios unchanged but reduce the numbers of participating monomers in proportion, can we use either the Flory-Stockmayer theory or the MC simulations to generate a molecular weight distribution that is applicable to a macroscopic system? To answer this question, we test 4 additional systems listed in Table~\ref{table:casestb2}. The smallest system has 10 PA, 134 BPADA, 146 MPD, and 2 TAPE and is denoted as $S_1$. Then the numbers of monomers are increased 10, 50, and 80 fold by keeping the ratios to generate systems $S_{10}$, $S_{50}$, and $S_{80}$. The subscript of the system label reflects the size ratio with respect to the smallest system, $S_1$. In this notation, the system $S_<$ in Table~\ref{table:casestb1} is equivalent to $S_5$. All these systems are still below the gel point when fully reacted.

\begin{table*}[h]
\resizebox{\textwidth}{!}{%
\begin{tabular}{|c|c|c|c|c|c|c|c|c|c|c|c|c|}
\hline
       & PA   & BPADA & MPD & TAPE & $\rho_1$ & $\rho_2$ & $p_A$ & $p_B$ & $\alpha$ & $M_n$ (Da) & $M_w$ (Da)  & $M_z$ (Da)  \\ \hline
$S_{1}$ & 10 & 134     & 136   & 2 &0.0360 & 0.0216 & 1 & 1 & 0.366 & $15742\pm 14$ & $30829\pm 46$ &$37334\pm 57$ \\ \hline
$S_{10}$ & 100 & 1340     & 1360   & 20 &0.0360 & 0.0216 & 1 & 1  & 0.366 & $19799\pm 18$ & $59940\pm 607$ & $101321\pm 1441$ \\ \hline
$S_{50}$ & 500 & 6700     & 6800   & 100 &0.0360 & 0.0216 & 1 & 1  & 0.366 & $20361\pm 4$ & $73582\pm 619$ & $161904\pm 2612$ \\ \hline
$S_{80}$ & 800  & 10720   & 10880 & 160 &0.0360 & 0.0216 & 1 & 1  & 0.366 & $20417\pm 3$ & $75980\pm 550$ & $177919\pm 2712$\\ \hline
\end{tabular}}
\caption{Four fully reacted, stoichiometric systems all below the gel point but with the size increased proportionally without changing the fraction of each type of monomers. The entries have the same format as in Table~\ref{table:casestb1}. The subscript of the system label in the first column indicates the size ratio with respect to the base system, $S_1$. The average molecular weights, $M_n$, $M_w$, and $M_z$, are from the MC simulations.}
\label{table:casestb2}
\end{table*}

\begin{figure}
\center
\includegraphics[width = 0.45\textwidth]{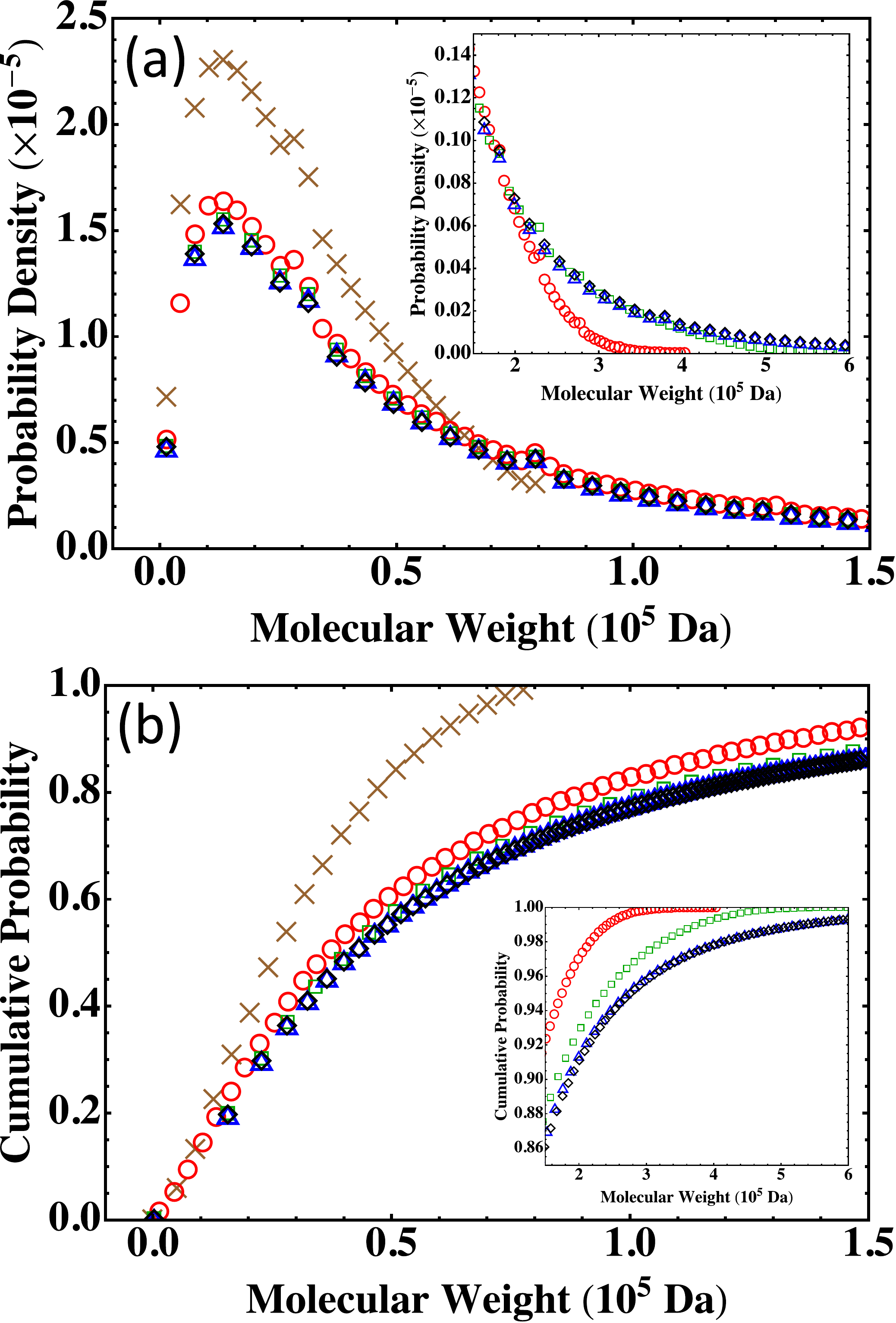}
\caption{Molecular weight distribution predicted by the Flory-Stockmayer theory for systems with different sizes: (a) Probability density and (b) cumulative probability. The main panels show the data in the low molecular weight region while the insets show the data in the high molecular weight region. Data are for $S_1$ (brown crosses), $S_<$ (red circles), $S_{10}$ (green squares), $S_{50}$ (blue triangles), and $S_{80}$ (black diamonds).}
\label{fig:system_size}
\end{figure}

The molecular weight distributions predicted by the Flory-Stockmayer theory for $S_1$, $S_<$, $S_{10}$, $S_{50}$, and $S_{80}$, including the probability density and the cumulative probability, are shown in Fig.~\ref{fig:system_size}. The main panels are for the region of low molecular weight and the insets show the data in the high molecular weight region. The data show that when the system size is increased, the curves of the molecular weight distribution converge quickly. There is a clear difference between the data for $S_1$ and those for $S_<$ (i.e., $S_5$). However, the difference between $S_<$ and $S_{80}$ is very small in the low molecular weight region and only discernible in the tail of the distribution in the region of high molecular weight (see the insets of Fig.~\ref{fig:system_size}). Furthermore, the results for $S_{50}$ and $S_{80}$ are almost indistinguishable in the entire region of molecular weight relevant to experiments, indicating that these systems are already large enough such that the molecular weight distribution is not affected by the finite system size any more.

\begin{figure}
\center
\includegraphics[width = 0.45\textwidth]{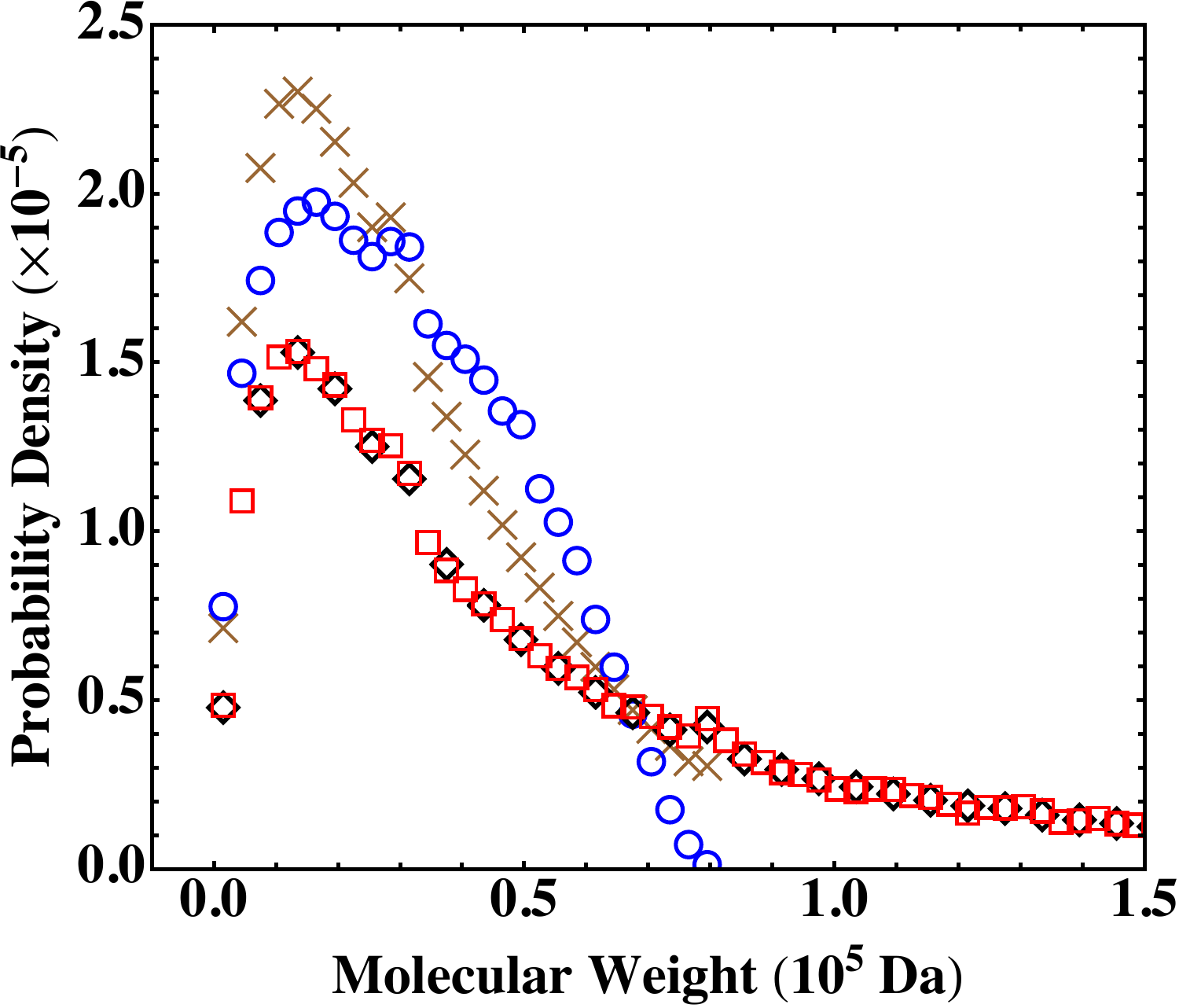}
\caption{Molecular weight distribution with data from for the Flory-Stockmayer theory for $S_1$ (brown crosses), the MC simulations for $S_1$ (blue circles), the Flory-Stockmayer theory for $S_{80}$ (black diamonds), and the MC simulations for $S_{80}$ (red squares). The MC results are averages of 50000 runs for $S_1$ and 1000 runs for $S_{80}$ (as well as $S_<$, $S_{10}$, and $S_{50}$).}
\label{fig:stockmayer_mc_S1_S80}
\end{figure}

Since $S_1$, $S_<$, $S_{10}$, $S_{50}$, and $S_{80}$ are all below the gel point, we expect the results from the Flory-Stockmayer theory and the MC simulations on the molecular weight distribution to agree. The comparison between the two is shown in Fig.~\ref{fig:stockmayer_mc_S1_S80} for $S_1$ and $S_{80}$. For $S_1$, some difference is observed between the prediction of the Flory-Stockmayer theory and the MC result because of the small size of this system. An excellent agreement is found between the theory and simulations for $S_{80}$. Similar agreements are also found for $S_{10}$ and $S_{50}$. A good agreement is already discussed earlier for $S_<$ as shown in Fig.~\ref{fig:three_systems}(a). These comparisons once again confirm that the Flory-Stockmayer theory provides a good description of the molecular weight distribution for systems well below the gel point, where ring formation is not a big concern. Below the gel point, both the Flory-Stockmayer theory and the MC simulations can be applied to a system containing only a few hundred to a few thousand monomers but having the same molar ratios of monomers as a macroscopic system to accurately predict the molecular weight distribution. As discussed earlier, the Flory-Stockmayer theory starts to fail when a system approaches or goes above the gel point. However, in these situations the MC simulations can still be used to quickly generate a  molecular weight distribution that is applicable to an experimental system.

\subsection{Partially Reacted Stoichiometric Systems} \label{sec:nfr_sys}

Up to this point, we mainly focus on fully reacted stoichiometric systems as it is possible to compute the molecular weight distribution using the Stockmayer formula even for a system with a relatively large size such as $S_{80}$. In this and next section we show that the conclusions reached so far also apply to partially reacted and/or nonstoichiometric systems. However, because of the practical difficulty of using the Stockmayer formula to compute the molecular weight distribution when either $p_A$ or $p_B$, or both, are less than 1, we use small systems with sizes similar to $S_5$ to illustrate the main point.

In this section we discuss partially reacted stoichiometric systems where $\sum_{q=1}^i f_qA_q=\sum_{h=1}^j g_hB_h$ but $p_A = p_B < 1$. Five such systems with the same size as $S_>$ are listed in Table~\ref{table:casestb3} where the values of $p_A$ and $p_B$ are increased from 0.95 to 0.99. The corresponding values of $\alpha$ changes from about 0.42 to about 0.65, thus enclosing the gelation transition at $\alpha_c = 0.5$.

\begin{table*}[h]
\resizebox{\textwidth}{!}{%
\begin{tabular}{|c|c|c|c|c|c|c|c|c|c|c|c|c|}
\hline
      & PA & BPADA & MPD & TAPE & $\rho_1$ &$\rho_2$ & $p_A$ & $p_B$ & $\alpha$ & $M_n$ (Da) & $M_w$ (Da)  & $M_z$ (Da)  \\ \hline
$S^{0.95}$ & 50 & 670     & 620   & 50 &0.0360 & 0.108 & 0.95 & 0.95 & 0.419 & $5965 \pm 3$ & $24779\pm 331$ & $47952\pm 824$ \\ \hline
$S^{0.96}$ & 50 & 670     & 620   & 50 &0.0360 & 0.108 & 0.96 & 0.96 & 0.462 & $7386\pm 5$ & $36488\pm 488$ & $69368\pm 1097$ \\ \hline
$S^{0.97}$ & 50 & 670     & 620   & 50 &0.0360 & 0.108 & 0.97 & 0.97 & 0.513 & $9655\pm 10$ & $58237\pm 807$ & $103512\pm 1504$\\ \hline
$S^{0.98}$ & 50 & 670     & 620   & 50 &0.0360 & 0.108 & 0.98 & 0.98 & 0.574 & $13780\pm 23$ & $105576\pm 1435$ & $165690\pm 2080$  \\ \hline
$S^{0.99}$ & 50 & 670     & 620   & 50 &0.0360 & 0.108 & 0.99 & 0.99 & 0.649 & $22710\pm 70$ & $200514\pm 1915$ & $266624\pm 2083$ \\ \hline
\end{tabular}}
\caption{Five partially reacted, stiochiometric systems (i.e., $p_A = p_B < 1$). The entries have the same format as in Table~\ref{table:casestb1}. The superscript of the system label indicates the values of $p_A$ and $p_B$. The first two are below and the rest three are beyond the gel point. The average molecular weights, $M_n$, $M_w$, and $M_z$, are from the MC simulations.}
\label{table:casestb3}
\end{table*}

\begin{figure}[h]
\center
\includegraphics[width = 0.45\textwidth]{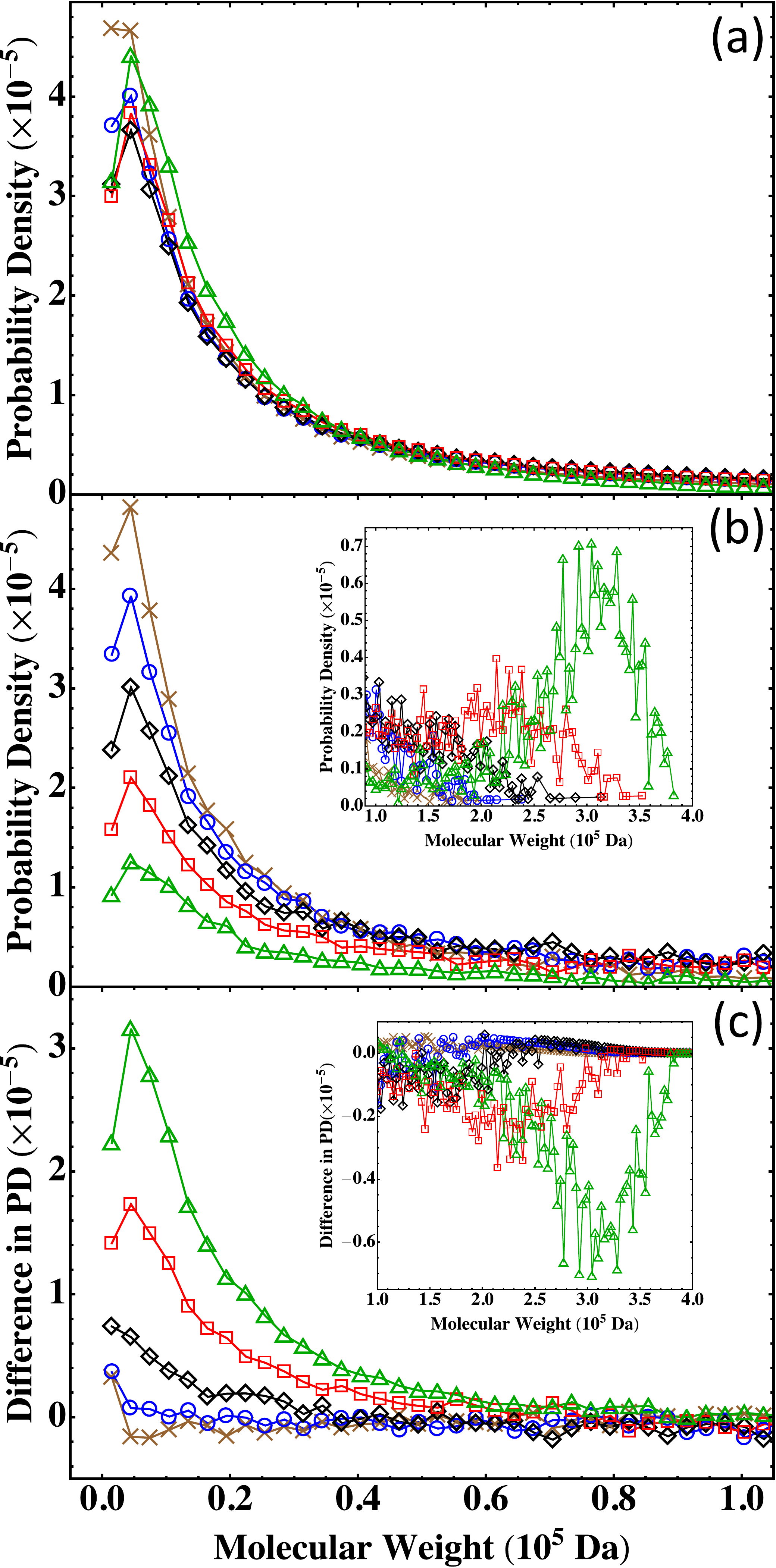}
\caption{Molecular weight distribution (a) from the Flory-Stockmayer theory and (b) from the MC simulations. The inset of (b) shows the MC results in the high molecular weight region. (c) Difference between the results from the Flory-Stockmayer theory and the MC simulations on the probability density (PD). The inset of (c) shows the difference in the high molecular weight region. The data are for $S^{0.95}$ (brown crosses), $S^{0.96}$ (blue circles), $S^{0.97}$ (black diamonds), $S^{0.98}$ (red squares), and $S^{0.99}$ (green triangles). The MC results are averages of 1000 runs.}
\label{fig:stoichiometric_pa95_pa99}
\end{figure}

The results on the molecular weight distribution from the Flory-Stockmayer theory and MC simulations at various values of $p_A$ and $p_B$ are shown in Figs.~\ref{fig:stoichiometric_pa95_pa99}(a) and (b), respectively. The molecular weight distribution predicted by the Flory-Stockmayer theory seems to be relatively insensitive to the values of $p_A$ and $p_B$. However, the MC results show that when the value of $p_A$ and $p_B$ is increased, the probability density in the low molecular weight region is reduced (see Fig.~\ref{fig:stoichiometric_pa95_pa99}(b)) while that in the high molecular weight region is enhanced (see the inset of Fig.~\ref{fig:stoichiometric_pa95_pa99}(b)). This systematic trend is expected as when the extent of reaction is larger, more polymers with higher molecular weights are anticipated to form.

To compare the predictions of the Flory-Stockmayer theory to the MC results on the molecular weight distribution, in Fig.~\ref{fig:stoichiometric_pa95_pa99}(c) their differences are shown for various $p_A$ and $p_B$. It is clear that when $p_A$ and $p_B$ are small, the systems are below the gel point and the results from the theory and simulations agree, as for $S^{0.95}$ and $S^{0.96}$. The difference becomes noticeable when the system approaches the gel point, such as $S^{0.97}$. For $S^{0.98}$ and $S^{0.99}$, they are above the gel point and clear differences in the probability density from the theory and simulations can be noted in both low (see Fig.~\ref{fig:stoichiometric_pa95_pa99}(c)) and high (see the inset of Fig.~\ref{fig:stoichiometric_pa95_pa99}(c)) molecular weight regions. The results for the partially reacted stoichiometric systems thus corroborate the conclusion that the Flory-Stockmayer theory only applies to systems well below the gel point. However, the MC simulations can be used to compute the molecular weight distribution for any systems no matter they are below, around, or above the gel point.

\subsection{Nonstoichiometric Systems} \label{sec:neq_sys}

We finally discuss nonstoichiometric systems where $\sum_{q=1}^i f_qA_q \neq \sum_{h=1}^j g_hB_h$ and $p_A \neq p_B$. Three systems with sizes similar to $S_>$ are shown in Table~\ref{table:casestb4}. We fix the value of $p_A$ at 0.99 but vary $p_B$ from 0.93 to 0.97. The numbers of monomers of PA, BPADA, and TAPE are all fixed. The number of MPD is varied according to Eq.~(\ref{eq:constraint}). Specifically, when the number of MPD is reduced, the values of $p_B$, $\rho_2$, and $\alpha$ are all increased. For the three systems in Table~\ref{table:casestb4}, $S_{<}^{n}$ is below, $S_{\simeq}^{n}$ is around, and $S_{>}^{n}$ is above the gel point. Here the superscript $n$ in the system labels indicates that these systems are nonstoichiometric.

\begin{table*}[h]
\resizebox{\textwidth}{!}{%
\begin{tabular}{|c|c|c|c|c|c|c|c|c|c|c|c|c|}
\hline
      & PA & BPADA & MPD & TAPE & $\rho_1$ & $\rho_2$ & $p_A$ & $p_B$ & $\alpha$ & $M_n$ (Da) & $M_w$ (Da)  & $M_z$ (Da)   \\ \hline
$S_{<}^{n}$ & 50 & 670   & 664 & 50   & 0.0360 & 0.101  & 0.99 & 0.93 & 0.445 & $7225 \pm 2$ & $33130\pm 195$ & $62388\pm 448$  \\ \hline
$S_{\simeq}^{n}$ & 50 & 670   & 649 & 50   & 0.0360  & 0.104 & 0.99 & 0.95 & 0.502 & $9530\pm 4$ & $54269\pm 349$ & $96424\pm 656$ \\ \hline
$S_{>}^{n}$ & 50 & 670   & 634 & 50   & 0.0360  & 0.106  & 0.99 & 0.97 & 0.569 & $13824\pm 10$ & $103074\pm 624$ & $162837\pm 915$\\ \hline
\end{tabular}}
\caption{Three partially reacted, nonstiochiometric systems (i.e., $p_A \neq p_B$ and both are less than 1) below, around, and beyond the gel point. The entries have the same format as in Table~\ref{table:casestb1}. The average molecular weights, $M_n$, $M_w$, and $M_z$, are from the MC simulations.}
\label{table:casestb4}
\end{table*}

\begin{figure}[h]
\center
\includegraphics[width = 0.4\textwidth]{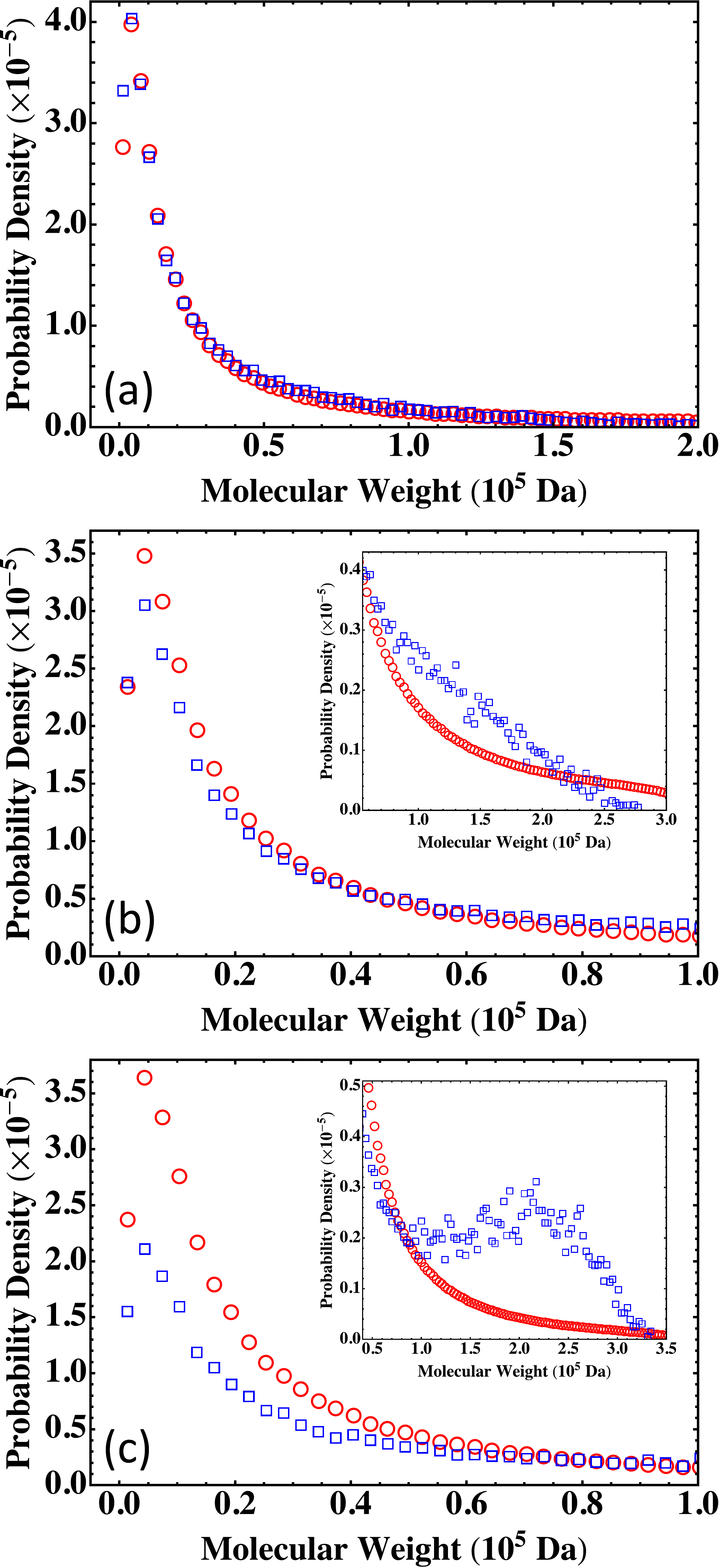}
\caption{Molecular weight distribution for the three systems in Table~\ref{table:casestb4}: (a) $S_<^{n}$, (b) $S_\simeq^{n}$, and (c) $S_>^{n}$. The results are for the Flory-Stockmayer theory (red circles) and the MC simulations (blue squares). The MC results are averages of 5000 runs.}
\label{fig:nonstoi_three_systems}
\end{figure}

The results on the molecular weight distribution for the three nonstoichiometric systems are plotted in Fig.~\ref{fig:nonstoi_three_systems}. For $S_{<}^{n}$ which is below the gel point, the MC results agree with the prediction of the Flory-Stockmayer theory, as shown in Fig.~\ref{fig:nonstoi_three_systems}(a). The two start to differ when a system approaches the gel point. An example is shown in Fig.~\ref{fig:nonstoi_three_systems}(b) for $S_{\simeq}^{n}$ with $\alpha = 0.502$. For this system, the Flory-Stockmayer theory overestimates the probability of low molecular weight polymers while underestimates the probability in the region of molecular weight higher than about $0.5\times 10^5$ Da (see the inset of Fig.~\ref{fig:nonstoi_three_systems}(b)). For $S_{>}^{n}$ which is above the gel point, the MC results on the probability density are smaller than those calculated with the Flory-Stockmayer theory when the molecular weight is lower than about $1\times 10^5$ Da (Fig.~\ref{fig:nonstoi_three_systems}(c)) but higher at higher molecular weights (see the inset of Fig.~\ref{fig:nonstoi_three_systems}(c)). For $S_{>}^{n}$ the molecular weight distribution has a second peak around $2\times 10^5$ Da, while the Flory-Stockmayer theory predicts a monotonically decaying distribution in this region. The results on nonstoichiometric systems thus one more time indicate that the Flory-Stockmayer theory only applies to systems well below the gel point, for which the formation of cyclic polymers or closed loops is negligible.

\section{Conclusions}
\label{sec:conclusions}

We have used MC simulations to study the polymerization of branched PEIs from BPADA (backbone monomer), MPD (backbone monomer), PA (chain terminator), and TAPE (branching agent). All the reactions for this system can be reduced to a condensation reaction between an amine group and a carboxylic anhydride group and thus can be characterized by one reaction rate. Our work show that in the MC model, the reaction rate should be computed using the concentrations of the functional groups on the monomers involved in a specific reaction, not the concentrations of the monomers themselves. The MC results are compared to the predictions of the Flory-Stockmayer theory. A practical approach of using the Flory-Stockmayer theory to compute molecular weight distributions has been suggested. We find that both the Flory-Stockmayer theory and the MC simulations accurately predict the molecular weight distribution for systems well below the gel point that is set by the functionality of the branching agent, though ring formation is not considered by the Flory-Stockmayer theory but allowed in MC simulations. The agreement between the theory and simulations thus indicates that ring formation is negligible for systems well below the gel point. However, for systems close to or above the gel point, the Flory-Stockmayer theory is not applicable as many cyclic polymers can be produced and ring structures can form in highly branched networks. For these systems, the MC simulations can still be used to quickly compute the molecular weight distribution that can be used to describe experimental measurements including average molecular weights.

Our tests indicate that in the MC simulations, a system with only a few hundred to a few thousand monomers but the same molar ratios of participating monomers is large enough to yield converging results on the molecular weight distribution for the region of molecular weight relevant to typical experiments (from 0 to about $3\times 10^5$ Da in the case of PEIs). These conclusions have been thoroughly confirmed with simulations for fully reacted, partially reacted, stoichiometric, and nonstoichiometric systems. The MC model presented here is expected to be applicable to a wide range of step-growth polymers.

\section*{Acknowledgement}
This paper is based on the results from work supported by SABIC Innovative Plastics US LLC.


\end{document}